\documentclass[numberedappendix]{emulateapj}

\usepackage{epsfig}
\usepackage{verbatim}
\usepackage{dcolumn}
\usepackage{graphicx}
\usepackage{bm}
\usepackage{epstopdf}
\usepackage{rotating}
\usepackage{url}
\usepackage{amssymb}
\usepackage{amsmath}
\usepackage{amsfonts}

\usepackage{xcolor}
\definecolor{darkgreen}{rgb}{0.0,0.5,0.0}
\usepackage[colorlinks,linkcolor=darkgreen,citecolor=darkgreen]{hyperref} 

\newcommand{\eg}{\emph{e.g.,} }

\newcommand{\be}{\begin{equation}}
\newcommand{\ee}{\end{equation}}
\newcommand{\bea}{\begin{equation*}}
\newcommand{\eea}{\end{equation*}}
\newcommand{\beqr}{\begin{eqnarray} \nonumber}
\newcommand{\eeqr}{\end{eqnarray}}
\newcommand{\beqrb}{\begin{eqnarray}}
\newcommand{\eeqrb}{\nonumber \end{eqnarray}}
\newcommand{\fin}{\mbox{ .}}
\newcommand{\coma}{\mbox{ ,}}

\newcommand{\const}{\mbox{const.}}

\newcommand{\grad}{\bm{\nabla}}
\newcommand{\vect}[1]{\bm{#1}}
\newcommand{\unit}[1]{\bm{\hat{#1}}}
\newcommand{\pr}{\partial}

\newcommand{\myNi}{\emph{(i)}\,}
\newcommand{\myNii}{\emph{(ii)}\,}
\newcommand{\myNiii}{\emph{(iii)}\,}
\newcommand{\myNiv}{\emph{(iv)}\,}

\def\myfig#1{./#1}

\newcommand{\myS}{{S}}
\newcommand{\myW}{{W}}
\newcommand{\overbar}[1]{\mkern 1.5mu\overline{\mkern-1.5mu#1\mkern-1.5mu}\mkern 1.5mu}
\newcommand{\mybar}[1]{\overbar{#1}}

\newcommand{\overtilde}[1]{\mkern 1.5mu\widetilde{\mkern-1.5mu#1\mkern-1.5mu}\mkern 1.5mu}
\newcommand{\mytilde}[1]{\tilde{#1}}
\newcommand{\myttilde}[1]{\overtilde{#1}}
\newcommand{\mystag}[1]{{#1}_0}
\newcommand{\myGamma}{\gamma}   
\newcommand{\mycs}{c}           
\newcommand{\mycsStag}{\bar{c}} 
\newcommand{\mySO}{\Delta}      
\newcommand{\myg}{g}            

\newcommand{\fixapj}[1]{{#1}}

\begin{document}

\title{Compressible flow in front of an axisymmetric blunt object:
\\ \fixapj{analytic approximation} and astrophysical implications}
\shorttitle{Compressible flow}
\shortauthors{Keshet \& Naor}
\author{Uri Keshet}
\author{Yossi Naor}

\affil{Physics Department, Ben-Gurion University of the Negev, PO Box 653, Be'er-Sheva 84105, Israel; ukeshet@bgu.ac.il}

\begin{abstract}
Compressible flows around blunt objects have diverse applications, but current analytic treatments are inaccurate and limited to narrow parameter regimes.
We show that the gas-dynamic flow in front of an axisymmetric blunt body is accurately derived analytically using a low order expansion of the perpendicular gradients in terms of the parallel velocity.
This reproduces both subsonic and supersonic flows measured and simulated for a sphere, including the transonic regime and the bow shock properties.
Some astrophysical implications are outlined, in particular for planets in the solar wind and for clumps and bubbles in the intergalactic medium.
The bow shock standoff distance normalized by the obstacle curvature is $\sim 2/(3g)$ in the strong shock limit, where $g$ is the compression ratio.
For a subsonic Mach number $M$ approaching unity, the thickness $\delta$ of \fixapj{an initially weak,} draped magnetic layer is a few times larger than in the incompressible limit, with amplification $\sim ({1+1.3M^{2.6}})/({3\delta})$.
\end{abstract}


\maketitle

\section{Introduction}

Compressible flows around blunt objects play an important role in diverse fields of science and engineering, ranging from fluid mechanics \citep[\eg][]{LandauLifshitz59_FluidMechanics, ParkEtAl2006,MackSchmid2011,TuttyEtAl2013,GrandemangeEtAl2013,GrandemangeEtAl2014}, space physics \citep[\eg][]{SpreiterAlksne70, BaranovLebedev88, SpreiterStahara95, CairnsGrabbe94, PetrinecRussell97, Petrinec02}, and astrophysics \citep{LeaYoung76, ShavivSalpeter82, CantoRaga98, SchulreichBreitschwerdt11}, to computational physics and applied mathematics \citep{Hejranfaretal09, Wilson13, GollanJacobs13, Marroneetal13}, aeronautical and civil engineering \citep{NAKANISHIKAMEMOTO93, Baker10, Aulchenkoetal12}, and aerodynamics \citep{AsanalievEtAl88,LiouTakayama05,PilyuginKhlebnikov06,Volkov09}.
Yet, even for the simple case of an inviscid flow around a sphere, the problem has resisted a general or accurate analytic treatment due to its nonlinear nature.

In particular, in space physics and astrophysics, the interaction of an ambient medium with much denser, in comparison approximately solid, bodies such as comets \citep[\eg][]{BaranovLebedev88}, planets \citep{SpreiterAlksne70,CairnsGrabbe94,PetrinecRussell97}, binary companions \citep{CantoRaga98}, galaxies \citep{ShavivSalpeter82,SchulreichBreitschwerdt11}, or large scale clumps and bubbles \citep{LeaYoung76, Vikhlininetal01, Lyutikov06, MarkevitchVikhlinin07}, is important for modeling these systems and understanding their observational signature. This is particularly true for the shocks formed in supersonic flows, due to their rich nonthermal effects \citep[\eg][]{SpreiterStahara95, Vikhlininetal01, Petrinec02, MarkevitchVikhlinin07}.

Although these fairly complicated systems can be approximately solved numerically, they are often modeled as an idealized, inviscid flow around a simple blunt object, often approximated as axisymmetric or even spherical, with some simplified analytic description employed in order to gain a deeper, more general understanding of the system.
Consequently, this fundamental problem of fluid mechanics has received considerable attention.
The small Mach number $M$ regime was studied as an asymptotic series about $M=0$ \citep{LordRayleigh1916, tamada39, kaplan40,StangebyAllen71,Allen13}, and solved in the incompressible potential flow limit.
Some hodograph plane results and series approximations were found in the transonic and supersonic cases \citep{Hida1955asymptotic, LiepmannRoshko57, Guderley_TransonicFlow}.
In particular, approximations for the standoff distance of the bow shock \citep[\eg][]{Moeckel49, Hida53, Lighthill57, HayesProbstein66, SpreiterEtAl66, Guy74, CoronaRomero13} partly agree with experiments \citep[\eg][]{Heberle_etal50, SchwartzEckerman56}, spacecraft data \citep{FarrisRussell94, SpreiterStahara95, Veriginetal99}, and numerical computations \citep{ChapmanCairns03, IgraFalcovitz10}.

However, these analytic results are typically based on ad hoc, unjustified assumptions, such as negligible compressibility effects, a predetermined shock geometry \citep{Lighthill57, Guy74}, or an incompressible \citep{Hida53} or irrotational \citep{kawamura1950mem, Hida1955asymptotic} flow downstream of the shock.
Other approaches use slowly converging, or impractically complicated, expansion series \citep{LordRayleigh1916, Hida1955asymptotic, vanDyke58a, vanDyke75Book}.
In all cases, the results are inaccurate or limited to a narrow parameter regime.
A generic yet accurate analytic approach is needed.

We adopt the conventional assumptions of \myNi an ideal, polytropic gas with an adiabatic index $\myGamma$; \myNii negligible viscosity and heat conduction (ideal fluid); \myNiii a steady, laminar, non-relativistic flow; and \myNiv negligible electromagnetic fields.
Typically, these assumptions hold in front of the object, but break down behind it and in its close vicinity.
We thus analyze the flow ahead of the object.

While spatial series expansions and hodograph plane analyses, when employed separately, are of limited use \citep[for reviews, see][]{vanDyke58b,vanDyke75Book}, we find that their combination gives good results over the full parameter range.
In particular, we expand the axial flow in terms of the parallel velocity, rather than of distance.
This yields an accurate, \fixapj{fully} analytic description of the \fixapj{gas-dynamic} flow, in both subsonic and supersonic regimes, already in a second or third order expansion, as shown in Fig. \ref{Fig:AllFlows}.

After introducing the flow equations in \S\ref{sec:FlowEquations}, in particular along the axis of symmetry, we derive the expansion series for the subsonic regime in \S\ref{sec:SubsonicFlow}, and for the supersonic regime in \S\ref{sec:SupersonicFlow}.
Some astrophysical implications are demonstrated in \S\ref{sec:Astro}, in particular for planetary bow shocks and for clumps and bubbles in the intergalactic medium (IGM).
We begin the analysis with a sphere, and \fixapj{outline the generalization} for arbitrary blunt axisymmetric objects in \S\ref{sec:Discussion}, where the results are summarized and discussed.
For convenience, the full results are given explicitly in Appendix \S\ref{sec:Explicit}.

\section{Flow equations}
\label{sec:FlowEquations}

Under the above assumptions, the flow is governed by the stationary continuity, Euler, and energy equations,
\begin{equation} \label{eq:FlowEquations}
\grad \cdot (\rho\vect{v})=0 \, ; \,\,\,\, (\vect{v}\cdot \grad)\vect{v}=-\frac{\grad P}{\rho} \, ; \,\,\, \,
\vect{v}\cdot \grad \left( \frac{P}{\rho^\myGamma} \right)=0\, ,
\end{equation}
where $\vect{v}$, $P$ and $\rho$ are the velocity, pressure and mass density.
At a shock, downstream (subscript $d$) and upstream ($u$) quantities are related by the shock adiabat \citep[\eg][]{LandauLifshitz59_FluidMechanics},
\begin{equation} \label{eq:JumpConditions}
\!\frac{\rho_d}{\rho_u} = \frac{v_u}{v_d} = \frac{(\myGamma+1) M_u^2}{(\myGamma-1)M_u^2+2 } \, ; \,\,\, \,
\frac{P_d}{P_u} = \frac{2\myGamma M_u^2+1-\myGamma}{\myGamma+1},
\end{equation}
with $M\equiv v/\mycs$, and $\mycs=(\gamma P/\rho)^{1/2}$ being the sound speed.

Along streamlines, Bernoulli's equation implies that
\begin{equation} \label{eq:Bernoulli}
w + v^2/2  = \mybar{w} = \const \coma
\end{equation}
where $w=\myGamma P/[(\myGamma-1)\rho]$ is the enthalpy, and a bar denotes (henceforth) a putative stagnation ($v=0$) point.
The far incident flow is assumed to be uniform and unidirectional, so $\mybar{w}$ is the same constant for all streamlines.  Equation~(\ref{eq:Bernoulli}) remains valid across shocks, as $w+v^2/2$ is the ratio between the normal fluxes of energy and of mass, each conserved separately across a shock. 

\begin{figure*}
\centerline{\hspace{3.5cm}\epsfxsize=23cm \epsfbox{\myfig{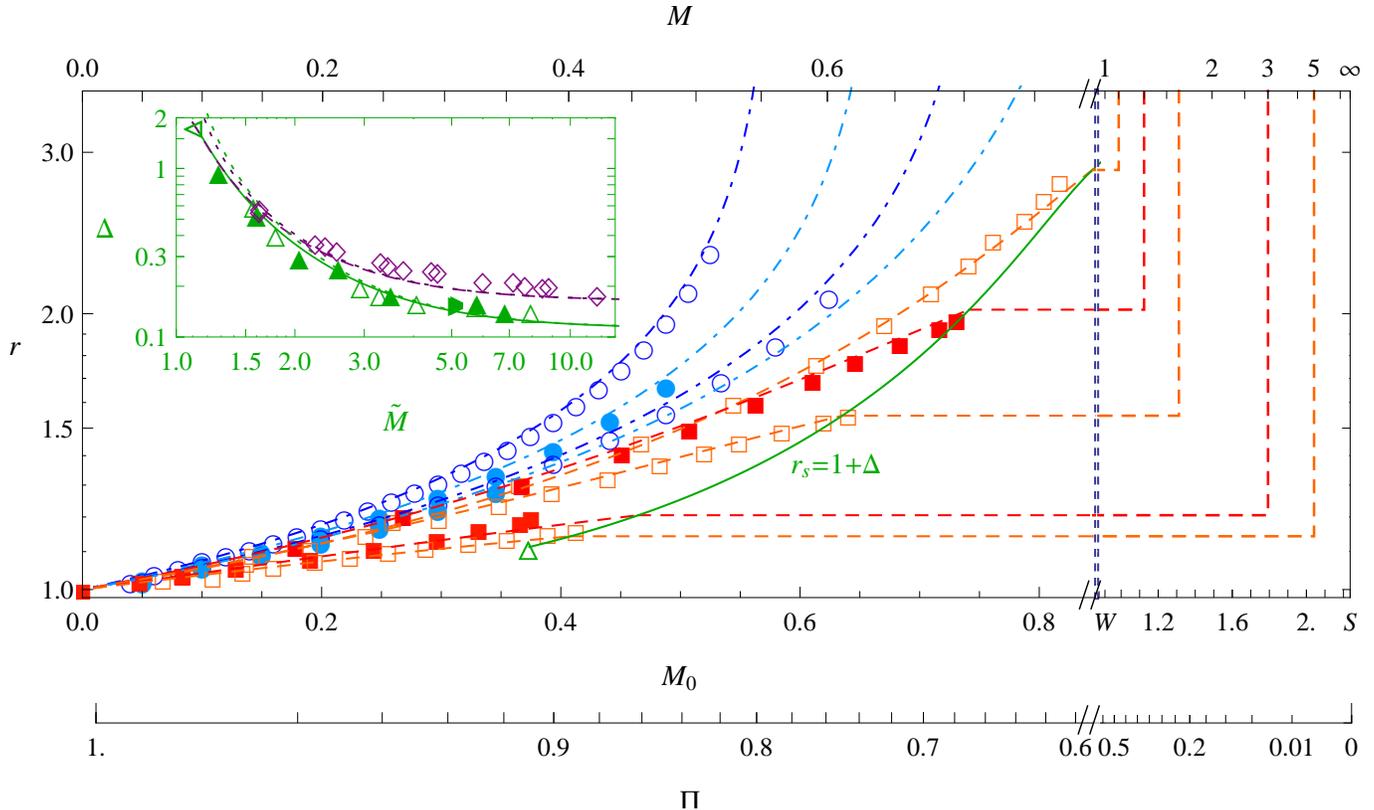}}}
\caption{
Radial profiles of Mach number $M$ (top axis) and of normalized velocity $M_0$ and pressure $\Pi$ (bottom axis; \fixapj{see definitions in Eq.~\ref{eq:M0andPi}}) in front of a unit ($r=1$) sphere, for $\myGamma=7/5$, according to numerical simulations (symbols) and our approximation (curves), in both subsonic (bluish circles and dot-dashed curves) and supersonic (reddish squares and dashed curves) regimes.
Numerical data shown (alternating shading to guide the eye) for $\myttilde{M}=0.6$, $0.7$, $0.8$, $0.95$ \citep{Karanjkar08}, $1.1$, $1.3$, $1.62$ \citep{Krause75, Heberle_etal50}, $3$ \citep{BonoAwruch08}, and $5$ \citep{Krause75, SedneyKahl61}.
The shock standoff distance (solid green) with its  $\myttilde{M}\to\infty$ limit (triangle) are also shown.
The right side of the figure extends it (on a different scale, to show the full $M$ range) to the supersonic, $M>1$ part of the flow, upstream of shocks; horizontal jumps represent the shock adiabat Eq.~(\ref{eq:JumpConditions}).
\emph{Inset}: standoff distance measured experimentally (symbols) and using the parameter-free (dotted curves; Eq.~(\ref{eq:xiSeries})) and single-parameter fit (Eq.~(\ref{eq:xiFit})) approximations, for $\myGamma=7/5$ (triangles; \citet{Heberle_etal50,vanDyke58b, SedneyKahl61, Krause75}; solid curve for $\beta=0.48$) and $\myGamma=5/3$ (diamonds; \citet{SchwartzEckerman56}; dashed curve for $\beta=0.52$).
\label{Fig:AllFlows}
}
\end{figure*}

Bernoulli's equation (\ref{eq:Bernoulli}) relates the local Mach number,
\begin{equation} \label{eq:M0andPi}
M 
= v/c = \left(\mystag{M}^{-2}-\myS^{-2}\right)^{-\frac{1}{2}} = ( \Pi^{-\frac{\myGamma-1}{\myGamma}}-1 )^{\frac{1}{2}} \myS \coma
\end{equation}
to the Mach number with respect to stagnation sound, $\mystag{M} \equiv v/\mycsStag$, and to the normalized pressure, $\Pi\equiv P/\mybar{P}$.
We define $\myS^2\equiv 2/(\myGamma-1)$ and $\myW^2\equiv 2/(\myGamma+1)$ as the strong and weak shock limits of $\mystag{M}^2$, so the subsonic (supersonic) regime becomes $0<\mystag{M}<\myW$ ($\myW<\mystag{M}<\myS$).
Figure~\ref{Fig:AllFlows} illustrates these definitions, and shows the shock adiabat Eq.~(\ref{eq:JumpConditions}) (as horizontal jumps at fixed $r$) for $\gamma=7/5$.

Consider the flow ahead of a sphere along the symmetry axis, $\theta=0$ in spherical coordinates $\{r,\theta,\phi\}$.
Here, the flow monotonically slows with decreasing $r$, down to $v=0$ at the stagnation point, which we normalize as $\mybar{\vect{r}}=\{1,0,0\}$.
Symmetry implies that along the axis $\vect{v}=-u(r)\unit{r}$, where $u>0$. Here, Eqs.~(\ref{eq:FlowEquations}) become
\begin{equation}
\label{eq:AxisEquations}
\frac{\pr\ln (\rho u)}{\pr\ln r^2}=\frac{q-u}{u}  \, ; \quad
\pr_r P=-\rho u\pr_r u \, ; \quad \pr_\theta P=0 \coma
\end{equation}
along with Bernoulli's Eq.~(\ref{eq:Bernoulli}), where we defined $q\equiv (\pr_\theta v_\theta)_{\theta=0}$ as a measure of the perpendicular velocity. Hence,
\begin{align} \label{eq:uODE}
\pr_r u & = \frac{2}{r}(q-u) \frac{1-\mystag{M}^2/\myS^2}{1-\mystag{M}^2/\myW^2} \fin
\end{align}

Our analysis relies on $u(r)$ being a monotonic function.
This allows us to write $q=q(u)$ as a function of $u$ and not of $r$.
Integrating Eq.~(\ref{eq:uODE}) thus yields
\begin{equation} \label{eq:uSolution}
2\ln r = \int_0^{u(r)} \frac{1-\mystag{M}(u')^2/\myW^2}{1-\mystag{M}(u')^2/\myS^2}
\,\, \frac{du'}{q(u')-u'} \, ,
\end{equation}
so given $q(u)$, the near-axis flow is directly determined.

Unlike $u(r)$, or other expansion parameters used previously, the $q(u)$ profile for typical bodies varies little, and nowhere vanishes.
It is well approximated by a few terms in a power expansion of the form
\begin{equation} \label{eq:qExpansion}
q(u) = q_0 + q_1 (u-U) + q_2(u-U)^2 + \ldots \coma
\end{equation}
where $U$ is a reference velocity,
so the integral in Eq.~(\ref{eq:uSolution}) can be analytically carried out to any order (see \S\ref{sec:Explicit}).
Moreover, we next show that the boundary conditions tightly fix $q(u)$,
giving a good approximation for the near axial flow.

First expand $q\simeq \mybar{q}$ near stagnation, with $U=\mybar{u}=0$.
An initially homogeneous subsonic or even mildly supersonic \citep{LandauLifshitz59_FluidMechanics} flow remains irrotational, $\grad\times \vect{v}=0$, in which case the lowest-order constraint is
\begin{equation} \label{eq:ConstPotentialFlow}
\mybar{q}_1 = -1/2 \coma
\end{equation}
whereas for a supersonic, rotational flow, it becomes
\begin{equation} \label{eq:ConstGeneralFlow}
3\mycsStag^2\mybar{q}_3+7\mycsStag\mybar{q}_2 = 2\mybar{q}_1 + 6\frac{\mybar{q}_0}{\mycsStag} + \mybar{q}_1\left(\frac{\mybar{q}_0}{\mycsStag}\right)^2 + \left(\frac{\mybar{q}_0}{\mycsStag}\right)^3 \coma
\end{equation}
as seen by expanding Eqs.~(\ref{eq:FlowEquations}) to order $\theta^2(r-1)^3$.
The generalization for non-spherical objects is discussed in \S\ref{sec:Discussion}.
Next, we estimate $q$ far from the body, and use it to approximate the flow in both the subsonic (\S\ref{sec:SubsonicFlow}) and supersonic (\S\ref{sec:SupersonicFlow}) regimes.

\section{Subsonic flow}
\label{sec:SubsonicFlow}

In the subsonic, $\tilde{M}<1$ case, we derive the incoming axial flow out to $r\to\infty$.
Using the incident flow (labeled by a tilde, henceforth) boundary condition $\mytilde{\vect{v}}=\mytilde{u}\{-\cos\theta,\sin\theta,0\}$, we may expand $\myttilde{q}$ with $U=\mytilde{u}$, such that
\begin{equation} \label{eq:ConstSubsonicQ0}
\myttilde{q}_0= (\pr_\theta \mytilde{v}_\theta)_{\theta=0} =\mytilde{u} \fin
\end{equation}
Additional terms can be derived using $\mytilde{M}\ll 1$ or $r\gg 1$ expansions appropriate for the relevant object.
Here, it will suffice to consider the leading, $(u-\mytilde{u})\propto r^{-\alpha}$ behavior at large radii, such that Eq.~(\ref{eq:uODE}) yields
\begin{equation} \label{eq:ConstSubsonicQ1}
\myttilde{q}_1 = 1 - \frac{\alpha}{2}\, \frac{1-\myttilde{M}_0^2/\myW^2}{1-\myttilde{M}_0^2/\myS^2} \fin
\end{equation}

In the incompressible limit, $\alpha=3$ for any object \citep[\eg][]{LandauLifshitz59_FluidMechanics}.
This also holds for general forward-backward symmetric objects in any potential flow.
To see the latter, expand the potential $\Phi$, defined by $\vect{v}=\tilde{u}\grad\Phi$, as a power series in $r$.
Imposing the $r\to\infty$ boundary conditions and regularity across $\theta=0$ yields
\begin{align}
\Phi = -r\cos\theta+\frac{\varphi_1}{r \Theta}+\frac{\varphi_2\cos\theta}{r^2 \Theta^3} + \ldots \coma
\end{align}
where $\Theta\equiv [1-M^2(\myS^{-2}+\sin^2\theta)]^{1/2}$.
The constants $\varphi_k$ are determined by the boundary conditions on the specific body.
Symmetry under forward-backward inversion, $\Phi\to-\Phi$ if $\theta\to \pi-\theta$, requires that $\varphi_1=0$.
In general $\varphi_2\neq 0$, implying that indeed $\alpha=3$.
Such behavior is demonstrated for an arbitrary compressible flow around a sphere by the Janzen-Rayleigh series \citep[\eg][]{tamada39,kaplan40}.

Finally, the $\myttilde{q}$ expansion at $r\to\infty$ is matched to the $\mybar{q}$ expansion at stagnation for a potential flow. In the limit of an incompressible flow around a sphere, Eqs.~(\ref{eq:ConstPotentialFlow}), (\ref{eq:ConstSubsonicQ0}), and (\ref{eq:ConstSubsonicQ1}) yield $q(u)=\tilde{u}-(u-\tilde{u})/2+O(u-\tilde{u})^2=3\tilde{u}/2-u/2$, which is indeed the exact solution.

This procedure reasonably approximates arbitrary compressible, subsonic flows. Better results are obtained by noting that the constraint (\ref{eq:ConstPotentialFlow}) holds also before stagnation, as long as $\pr_{\theta\theta}v_r$ is negligible, implying that $\mybar{q}_2\simeq 0$. Combining this with constraints~(\ref{eq:ConstPotentialFlow}), (\ref{eq:ConstSubsonicQ0}), and (\ref{eq:ConstSubsonicQ1}) yields an accurate, third order approximation, shown in Fig.~\ref{Fig:AllFlows} as dot-dashed curves.
See \S\ref{sec:ExplicitSubsonic} for an explicit solution.

\section{Supersonic flow}
\label{sec:SupersonicFlow}

In the supersonic, $\tilde{M}>1$ case, a detached bow shock forms in front of the object, at the so-called standoff distance $\mySO$ from its nose.
The transition between subsonic and supersonic regimes is continuous, so $\mySO\to\infty$ as $\myttilde{M}\to 1$, or equivalently as $\myttilde{M}_0\to \myW$.
The unperturbed upstream flow and the shock transition are shown on the right side of Fig.~\ref{Fig:AllFlows}.

Consider the flow between the shock and stagnation along the axis of symmetry.
The $q(u)$ profile is strongly constrained if the normalized shock curvature $\xi^{-1}\equiv (R/r_s)_{\theta=0}$ is known.
Here, $r_s$ is the shock radius, such that $r_s(\theta=0)=1+\mySO$, and $R=r_s/[1-r_s''(\theta)/r_s]$  is its local radius of curvature.

Expanding the flow Eqs.~(\ref{eq:FlowEquations}) using Eqs.~(\ref{eq:JumpConditions}) as boundary conditions, yields the $q^{(d)}$ expansion coefficients around $U=u_d$, just downstream of the shock,
\begin{equation} \label{eq:q0d}
q_{0}^{(d)} = \left(1+\myg \xi-\xi\right) \myg^{-1}\tilde{u} \, ;
\end{equation}
\begin{equation} \label{eq:q1d}
q_{1}^{(d)} = \frac{3+(\myg-3)\xi}{2}-\frac{1+(3\myg-1)\xi}{1+\myg+(\myg-1)\myGamma} \, ;
\end{equation}
and
\begin{flalign} \label{eq:q2d}
q_{2}^{(d)}  = & \frac{\myg\xi \myW^2}{8 \left(\myg+\myW^2-1\right)^2\myttilde{u}} \Big[\frac{ \myg^2-4 \myg+3}{\xi\myW^2} -\frac{ 2 (3\myg+1)}{\xi}  \\
 &+  2 \left(\myg^2+4 \myg+1\right) -\frac{(\myg-1)^2 (\myg+3)}{\myW^2}+\frac{8 \myg^2 \myW^2}{\myg-1} \Big] \coma  \nonumber
\end{flalign}
where $\myg\equiv (\myttilde{M}_0/\myW)^2\geq 1$ is the axial compression ratio.

These coefficients depend on the shock profile only through $\xi$; higher order terms are sensitive to deviations of the profile from a sphere of radius $R$.
In the weak shock limit $\myg\to 1$, so $\xi$ must vanish to avoid the divergence of $q_{2}^{(d)}$.
This implies that $R$ diverges faster than $\Delta$, and $q_{1}^{(d)}\to (1-2\xi)$ asymptotes to unity, consistent with a smooth transition to the subsonic regime.
Moreover, if we require that $q_2^{(d)}\to \tilde{q}_2 \to 3/(2\bar{c}W)$ in \fixapj{this} limit (see \S\ref{sec:ExplicitSubsonic}), then
\begin{equation} \label{eq:TransonicXi}
\xi(\myttilde{M}_0\simeq\myW) \to (4+\myGamma)(-1+\myttilde{M}_0/\myW) \coma
\end{equation}
so $R/r_s$ diverges as $(\myttilde{M}_0-\myW)^{-1}$, consistent with \citet[][as expected in the irrotational limit]{Hida53, Hida1955asymptotic}.

Equations~(\ref{eq:q0d})--(\ref{eq:q2d}) yield a good, second order approximation to the flow, as shown in Fig.~\ref{Fig:AllFlows} (dashed curves), once $\xi$ or any of the $q^{(d)}$ coefficients are determined.
This can be done using the stagnation boundary conditions, such as Eq.~(\ref{eq:ConstGeneralFlow}), but is laborious and body-specific due to the high order involved.
A simpler approach is to estimate $\xi(M)$ using the weak and strong shock limits.

In the strong shock, $\myttilde{M}_0\to\myS$ limit, the curvature of the shock approaches that of the object \citep[\eg][]{Guy74}; $\xi\to 1$ in the case of the sphere.
The $\myttilde{M}_0(\xi)$ relation may be derived as a power series, using this and the constraint Eq.~(\ref{eq:TransonicXi}).
A second order expansion in $\xi$ gives a good approximation, valid throughout the supersonic regime,
\begin{equation} \label{eq:xiSeries}
\frac{\myttilde{M}_0}{\myW}-1 \simeq \frac{\xi}{4+\myGamma} + \left(\frac{S}{W}-\frac{5+\myGamma}{4+\myGamma}\right) \xi^2 \coma
\end{equation}
with no free parameters. The good fit suggests that higher order terms in $\xi$ are negligible or absent.

Alternatively, the result $\xi(\myttilde{M}_0\to S)=1$ and direct measurements of $\xi$ \citep{Heberle_etal50}, motivate a power-law approximation of the form
\begin{equation} \label{eq:xiFit}
\xi \simeq \left[(\myttilde{M}_0-\myW)/(\myS-\myW)\right]^\beta \fin
\end{equation}
We find that Eq.~(\ref{eq:xiFit}) nicely fits the measured flow for Mach numbers not too small, with $\beta\simeq 1/2$.

The standoff distance may now be found by solving Eq.~(\ref{eq:uSolution}) for $r_s=1+\Delta$, taking $u=u_d$ or equivalently $M_0=\myttilde{M}_0/g=\myW^2/\myttilde{M}_0$, using the expansion (\ref{eq:qExpansion}) with coefficients (\ref{eq:q0d}--\ref{eq:q2d})  fixed by the $\xi(\myttilde{M}_0)$ relation.
The figure inset shows that Eq.~(\ref{eq:xiSeries}) provides a good fit to the standoff distance throughout the supersonic range, for two equations of state. It also shows that a single $\beta\simeq 1/2$ power-law in Eq.~(\ref{eq:xiFit}) reproduces $\Delta$ away from the transonic regime. Indeed, $\Delta$ is sensitive to the precise value of $\beta$ only in the $M\simeq 1$ limit; best results are obtained with $\beta=0.48$ ($\beta=0.52$) for $\gamma=7/5$ ($\gamma=5/3$).

\section{Astrophysical implications}
\label{sec:Astro}

The above prescription for the flow in front of a blunt object is useful in a wide range of astrophysical circumstances, as the low-density medium can often be approximated as ideal and inviscid, the body as impenetrable, and the motion as steady and non-relativistic.

\begin{figure}
\centerline{\epsfxsize=8.5cm \epsfbox{\myfig{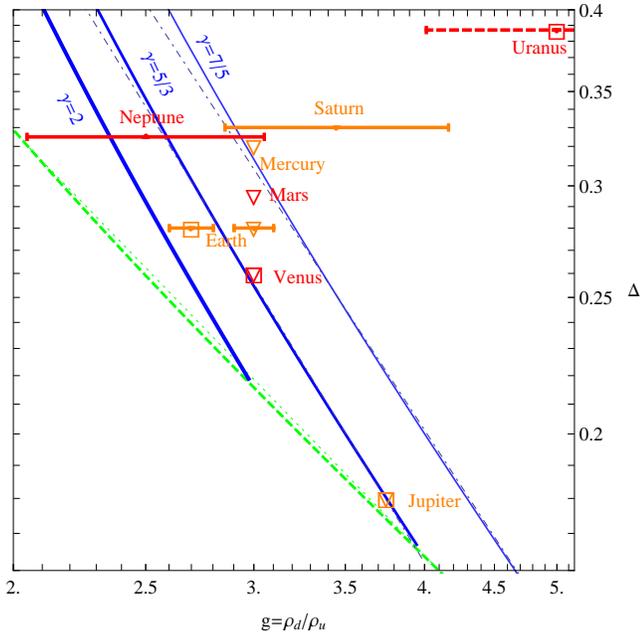}}}
\caption{
Bow shock standoff distance $\Delta$, measured from the nose of the obstacle and normalized by the nose curvature of the obstacle, plotted against the compression ratio $\myg$.
Analytic curves for $\gamma=7/5,\, 5/3$ and $2$ (solid, thin to thick) are shown alongside planet data (labeled symbols), and approximated as power laws (dot-dashed curves, see text).
Also plotted is the strong shock limit for various $\gamma$ (dashed), well fit by $\Delta\sim2/(3\myg)$ (dotted).
Planetary $g$ values are based on magnetic (triangle or no symbol) or density (square) compression, shown with $1\sigma$ error bars (when available) for
Mercury \citep[slightly perpendicular shock; measured at $\theta_p\simeq 70^\circ$;][]{AndersonEtAl08, TreumannJaroschek08, SlavinEtAl12},
Venus \citep[day side;][]{TreumannJaroschek08, FrankEtAl91},
Earth
\citep[quasi-perpendicular; $\theta_p\lesssim45^\circ$;][]{CzaykowskaEtAl00},
Mars \citep[quasi-perpendicular; $\theta_p\lesssim90^\circ$;][]{TreumannJaroschek08},
Jupiter \citep[$\theta_p\simeq 20^\circ$;][]{Gloeckler04},
Saturn \citep[quasi-perpendicular; interval for different crossings at high angles $60^\circ\lesssim\theta_p\lesssim100^\circ$;][figure 9]{AchilleosEtAl06},
Uranus \citep[$\theta_p\simeq 25^\circ$;][]{BagenalEtAl87},
and Neptune \citep[quasi-perpendicular; $\theta_p\simeq 14^\circ$;][]{NessEtAl89, TreumannJaroschek08}.
Standoff distances and obstacle curvatures are based on the data-constrained models of \citet[][for Venus, Earth and Mars]{Stahara1984} and \citet[][for the other planets]{SpreiterStahara95}.
Systematic errors on $\Delta$ are large, especially for the external and rarely visited planets; in particular, the $\Delta$ estimate for Uranus (dashed) is inconclusive \citep{SpreiterStahara95}. For details\fixapj{, assumptions and limitations}, see text.
\label{Fig:Planets}}
\end{figure}

Consider for example the standoff distance $\Delta$ in front of a supersonic astronomical object.
It is useful to plot $\Delta$ as a function of the compression ratio $g$, rather than of the Mach number, because it is typically easier to measure $g$.
As Fig. \ref{Fig:Planets} shows, $\Delta(g)$ at a given $\gamma$ approximately follows a power-law, for example $\Delta(\gamma=7/5)\simeq 1.6g^{-1.5}$ and $\Delta(\gamma=5/3)\simeq 1.5g^{-1.6}$.
For high $\tilde{M}$, the standoff distance approaches the strong shock limit, approximately given by \fixapj{(see \S\ref{sec:subsecSO})}
\begin{equation}
\Delta(\tilde{M}\to\infty) \simeq \frac{2}{3g} \fin
\end{equation}

For an arbitrary axisymmetric blunt body, the above results for a sphere are trivially generalized, if $\Delta$ is defined as the distance from the nose of the body, normalized by its radius of curvature (for additional corrections, see \S\ref{sec:Discussion}).
One may thus superimpose $\Delta(g)$ estimates of astronomical bow shocks on Fig. \ref{Fig:Planets}, even for non-spherical bodies.

Consider for example the bow shock of a planet, moving supersonically through the solar wind.
Although the magnetic Mach number and the ratio $\Delta/\lambda_i$ ($\lambda_i$ being the ion gyroradius) are not very high in such systems, a gas-dynamic approach remains useful as a first approximation, provided that $\tilde{M}$ is replaced by the fast magnetosonic Mach number upstream \citep{Stahara1984, SpreiterStahara95, FairfieldEtAl01}.
Here, we define $\Delta$ as the distance between the bow shock and the nose of the obstacle, namely the planetary magnetosphere or ionosphere, normalized by the radius of curvature of this obstacle's nose.
For a discussion of planetary bow shocks, and a compilation of $\Delta$ estimates based on analytic arguments and numerical simulations, see \citet[][and references therein]{VeriginEtAl03}.
Note that our analysis directly provides not only $\Delta(g)$, but also the flow profile and the shock radius of curvature.

Estimates of $\Delta(g)$ for the solar system planets are shown in Fig. \ref{Fig:Planets}, with references provided in the caption.
Interestingly, some planetary data seem to suggest a soft equation of state with $\gamma<5/3$.
However, such an interpretation is hindered by the substantial simplifying assumptions, in particular the neglected MHD effects, kinetic effects, variable solar wind conditions, and non-axisymmetric corrections to the obstacles.
The positions and shapes of the obstacles are in some cases highly uncertain; indeed, the results suggest a significant flattening of the magnetospheres of Saturn and Uranus.

As another astronomical system, consider the large scale extreme, namely the IGM of a galaxy group or cluster.
Here, hot bubbles inflated by the active galactic nucleus (AGN) rise buoyantly through the IGM \citep[\eg][]{FabianEtAl00,NulsenEtAl05}, and the subsonic motion of the plasma in front of them is important, for example, for computing the evolution of the bubbles \citep[\eg][]{ChurazovEtAl01_M87Bubble}, and the draping of magnetic fields around them \citep{Lyutikov06, DursiPfrommer08, NaorKeshet15}.
Large scale mergers lead to dense clumps moving subsonically or supersonically through the IGM, giving rise to dramatic effects such as shocks, cold fronts, and even a spatial separation between baryonic and dark matter components \citep{Vikhlininetal01, MarkevitchVikhlinin07}.
Details such as the bow shock location and the downstream flow pattern are important for correctly interpreting the underlying dynamics.

Consider first a supersonic clump moving through the AGN.
A well known example is the 1E0657-56, so-called bullet, cluster at redshift $z=0.296$, showing a merger nearly in the plane of the sky \citep{MarkevitchEtAl02_Bullet, BarrenaEtAl02_Bullet}.
The moving clump is seen as a bullet-shaped discontinuity, preceded by a bow shock with $g\simeq 3.0$ \citep{Markevitch06} and $\Delta\simeq 2.4\pm0.2$.
Our analysis indicates that the large $\Delta$ corresponds to a weak shock, with $\tilde{M}\simeq 1.1$ (for $\gamma\simeq 5/3$, used henceforth).
This is consistent with the $\sim 65^\circ$ asymptotic shock angle far from the nose, which also suggests a $\tilde{M}\simeq 1.1$ shock.
However, the high compression ratio corresponds to a much stronger shock with $\tilde{M}\simeq 3$, indicating that the system is not in a steady state.
Indeed, plotting the corresponding $\Delta(g)$ on Fig. \ref{Fig:Planets} would suggest an unrealistically soft equation of state.
Simulations indicate that the shock velocity can be higher by a factor of $1.7$ \citep{SpringelFarrar07} or even $6$ \citep{MilosavljevicEtAl07_Bulllet} than expected from the clump velocity, because the shock (i) moves faster than the clump; and (ii) plows through gas that is infalling towards the clump \citep{SpringelFarrar07}.
Evidently, Fig. \ref{Fig:Planets} provides a simple way to gauge the relaxation level of a system.

Next consider the subsonic IGM flow in front of an AGN bubble or a slow clump.
While previous studies \citep[\eg][]{Lyutikov06, DursiPfrommer08} have approximated the motion as incompressible, the inferred velocities are often nearly sonic \citep{ChurazovEtAl01_M87Bubble, MarkevitchVikhlinin07}, implying considerable compressibility effects.
To illustrate this, we compute the magnetization caused by the draping of a weak upstream magnetic field around the moving object.
\fixapj{The results are applicable only to weak magnetic fields, where Eqs.~(\ref{eq:FlowEquations}--\ref{eq:JumpConditions}) remain a good approximation.}

\begin{figure}
\centerline{\epsfxsize=8.5cm \epsfbox{\myfig{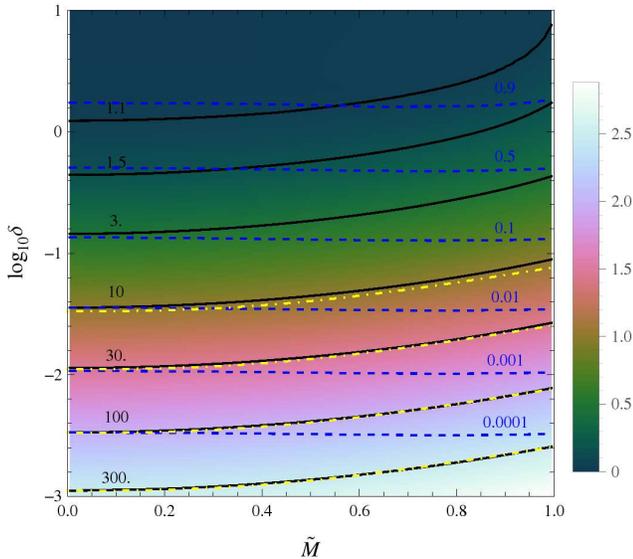}}}
\caption{
Energy amplification of a weak magnetic field initially perpendicular (solid contours, and $\log_{10}(B_{\perp}/\tilde{B}_{\perp})^2$ cube-helix \citep{Green11_Cubehelix} colormap) or parallel (dashed contours) to the subsonic $\gamma=5/3$ flow at a normalized distance $\delta=(r-1)$ in front of a sphere of Mach number $\tilde{M}$.
Close to the sphere, the field is predominantly perpendicular, and approximately given (dot-dashed contours) by Eq.~(\ref{eq:ApproxB}).
\label{Fig:MagPhaseSpace}
}
\end{figure}

The magnetic field generally evolves as $\bm{B}\propto \rho\bm{l}$, where $\bm{l}$ is a length element attached to the flow.
Hence, the magnetic components initially perpendicular or parallel to the flow evolve along the axis of symmetry according to
\begin{equation}
\frac{B_{\perp}}{\tilde{B}_{\perp}} = \left(\frac{\rho/v}{\tilde{\rho}/\tilde{v}}\right)^{1/2} = \left(\frac{M_0}{\tilde{M}_0} \right)^{-1/2} \left(\frac{S^2-M_0^2}{S^2-\tilde{M}_0^2}\right)^{S^2/4}
\end{equation}
or
\begin{equation}
\frac{B_{||}}{\tilde{B}_{||}} = \frac{\rho v}{\tilde{\rho} \tilde{v}} = \frac{M_0}{\tilde{M}_0}
\left( \frac{S^2-M_0^2}{S^2-\tilde{M}_0^2} \right)^{S^2/2} \, ;
\end{equation}
for a detailed discussion, see \citet{NaorKeshet15}.
The resulting magnetic energy amplification is shown in Fig. \ref{Fig:MagPhaseSpace}, for $\gamma=5/3$, as a function of $\tilde{M}$ and of the normalized distance from the body, $\delta=(r-1)$.
Near the object, the magnetization is predominantly perpendicular, and approximately given by
\begin{equation}\label{eq:ApproxB}
\frac{B_{\perp}}{\tilde{B}_{\perp}} \simeq \frac{1+1.3\tilde{M}^{2.6}}{3\delta} \coma
\end{equation}
as illustrated in the figure.

As the figure shows, the magnetized layer is typically a few times thicker for $\tilde{M}\simeq 1$ than it would appear in the incompressible limit.
Such thick layers may have observational implications, through their non-thermal pressure and as synchrotron emission in front of nearly sonic objects.
Such a synchrotron signal may contribute to the radio bright edges seen above AGN bubbles, for example in the Virgo cluster \citep{OwenEtAl00}.

\section{Discussion}
\label{sec:Discussion}

The compressible, inviscid flow in front of a blunt object is approximated analytically, using a hodograph-like, $\vect{v}\simeq (-u, q(u)\theta,0)$ transformation.
The velocity (Eq.~\ref{eq:uSolution}) and pressure (Eq.~\ref{eq:M0andPi}) profiles are derived by expanding $q$ as a (rapidly converging) power series in $u$ (Eq.~\ref{eq:qExpansion}), using the constraints imposed by the object (Eqs.~\ref{eq:ConstPotentialFlow} or \ref{eq:ConstGeneralFlow} for a sphere) and by the far upstream subsonic (Eqs.~\ref{eq:ConstSubsonicQ0}--\ref{eq:ConstSubsonicQ1}) or shocked supersonic (Eqs.~\ref{eq:q0d}--\ref{eq:q2d}) flow.
In the latter case, the weak (Eq.~\ref{eq:TransonicXi}) and strong shock limits approximately fix the shock curvature (Eq.~\ref{eq:xiSeries}) and consequently the flow, independent of the object shape.

Figure \ref{Fig:AllFlows} shows that a low order $q(u)$ expansion suffices to recover the measured flow in front of a sphere.
The supersonic results also reproduce the measured standoff distance (solid curve and figure inset) of the shock, and constrain its curvature (Eq.~\ref{eq:xiSeries} or the fit Eq.~\ref{eq:xiFit}).
Higher-order constraints can be used to improve the approximation further; here we used only the lowest-order constraint at stagnation, and only in the subsonic case.

The axial approximation directly constrains the flow beyond the axis and along the body, as it determines the perpendicular derivatives. For example, one can use it to estimate
$\pr_{\theta\theta}P = -\rho_0 [q^2-u\pr_r(rq)](1-\mystag{M}^2/S^2)^{1/(\gamma-1)}$, found by expanding Eqs.~(\ref{eq:FlowEquations}) to $\theta^2$ order.
Extrapolation beyond the axis is simpler in the potential flow regime, where, in particular, $\pr_{\theta\theta}v_r=\pr_r(rq)$.

The axial analysis is generalized for any blunt, axisymmetric object, by modifying the $q$ boundary conditions.
For a body with radius of curvature $R_b>0$ at a stagnation radius $r_b$, take $\{z\equiv r\cos\theta=R_b-r_b, \varrho\equiv r\sin\theta=0\}$ as the origin, and rescale lengths by $R_b$.
This maps the stagnation region of the body onto that of the unit sphere, so Eqs.~(\ref{eq:Bernoulli}--\ref{eq:ConstPotentialFlow}, \ref{eq:ConstSubsonicQ0}--\ref{eq:q2d}) remain valid.
The subsonic analysis is unchanged; for an asymmetric body, $\alpha$ may need to be altered, \eg using the Janzen-Rayleigh series.
The supersonic analysis is also unchanged, if Eq.~(\ref{eq:ConstGeneralFlow}) is used and adapted for the specific body.
The alternative use of Eq.~(\ref{eq:xiSeries}) or Eq.~(\ref{eq:xiFit}) is still expected to hold, although higher order terms or a tuned $\beta$ may be needed if an aspherical body modifies the weak or strong shock limits.

It may be possible to generalize our hodograph-like analysis even for a non-axisymmetric object, using the stagnant streamline instead of the symmetry axis, as long as the corresponding $u$ profile remains monotonic.

Our analysis is applicable to a wide range of subsonic and supersonic astronomical bodies.
Illustrative examples are discussed (in \S\ref{sec:Astro}), on both small, planetary scales, and large, galaxy cluster scales.
In particular, plotting the standoff distance as a function of the compression ratio (Fig. \ref{Fig:Planets}) can be used to gauge the equation of state and the relaxation level of the system.
The results are especially useful for nearly sonic flows, where compressiblity effects play an important role; this is seen for example in the thicker magnetically draped layers that form in front of a moving body (Fig. \ref{Fig:MagPhaseSpace}), such as a large scale clump or an AGN bubble.

\acknowledgements
We thank Ephim Golbraikh and Yuri Lyubarsky for helpful advice.
This research has received funding from the European Union Seventh Framework Programme (FP7/2007-2013) under grant agreement n\textordmasculine ~293975, from an IAEC-UPBC joint research foundation grant, from an ISF-UGC grant, and from an individual ISF grant.


\appendix

\section{Explicit description of the axial flow}
\label{sec:Explicit}

Here we present the flow in front of a sphere in explicit form, for both subsonic and supersonic regimes.
Fully analytic expressions are provided to second order in the supersonic case, and to third order in the subsonic case.
The generalization to an arbitrary axisymmetric blunt object is discussed in \S\ref{sec:Discussion}.

\subsection{Subsonic regime}
\label{sec:ExplicitSubsonic}

In the subsonic regime, $q(u)$ is constrained at stagnation and at infinity; we may expand it equivalently either around the stagnation point ($u=0$) or around infinity ($u=\myttilde{u}$).
Here we arbitrarily choose to write $q(u)$ using an expansion at stagnation, namely
\begin{equation} \label{eq:subsonic_q}
q = \mybar{q}_0-\frac{1}{2}u+\mybar{q}_3 u^3 + O(u^4)\coma
\end{equation}
where we used the stagnation boundary conditions $\bar{q}_1=(-1/2)$ and $\bar{q}_2\simeq 0$ derived in the main text.
This expression is matched with the boundary conditions at $r\to\infty$, in order to find the remaining coefficients,
\begin{equation}
\mybar{q}_0 = \frac{3}{2}\myttilde{u}-\mybar{q}_3 \myttilde{u}^3
\quad \mbox{and} \quad
\mybar{q}_3 = \frac{(S/W)^2-1}{2(S^2 \mybar{c}^2-\myttilde{u}^2)} \fin
\end{equation}

\subsection{Supersonic regime}
\label{sec:ExplicitSupersonic}

In the supersonic regime, we may write $q(u)$ using the downstream expansion at the shock,
\begin{equation} \label{eq:supsonic_q}
q=q_0^{(d)} + q_1^{(d)}(u-u_d)+q_2^{(d)}(u-u_d)^2 \coma
\end{equation}
where the coefficients are given in the main text.

\subsection{Explicit analytic results}

By plugging the relevant expressions for $q(u)$ into Eq.~(\ref{eq:uSolution}), $r(u)$ can be computed in both subsonic and supersonic cases.
In the latter, the shock radius is then determined as $r_s=1+\Delta = r(u_d)$.
The integral in Eq.~(\ref{eq:uSolution}) is easily evaluated numerically, using either Eq.~(\ref{eq:subsonic_q}) or (\ref{eq:supsonic_q}) for $q$.
However, the integral can also be carried out analytically, as follows.

\subsubsection{Subsonic regime}

In the subsonic case, plugging $q(u)$ from Eq.~(\ref{eq:subsonic_q}) into Eq.~(\ref{eq:uSolution}), and carrying out the integral, yields
\begin{align}
\ln(r)= \frac{S^2}{2 W^2} \, \frac{\bar{c} \left(S^2-W^2\right) \left(\frac{1}{2} \bar{c}
\left(\frac{3}{2}- S^2 \bar{c}^2 \bar{q}_3\right) \ln \left(1-\frac{u^2}{S^2 \bar{c}^2}\right)-\frac{\bar{q}_0}{S} \coth ^{-1}\left(\frac{S
   \bar{c}}{u}\right)\right)-F(u)+F(0)}{ \bar{q}_0^2-S^2
   \bar{c}^2 \left(S^2 \bar{c}^2 \bar{q}_3-\frac{3}{2}\right){}^2} \coma
\end{align}
where we defined
\begin{align}
F(u) \equiv H\Bigg[ &
x^3 \bar{q}_3+\bar{q}_0-\frac{3 x}{2}, \\
& \frac{\ln (u-x) \left(\bar{c}^2 \left(2 S^2 \bar{c}^2 \bar{q}_3-3\right) \left(W^2 \left(2
   \bar{q}_3 \left(S^2 \bar{c}^2+x^2\right)-3\right)-2 S^2 x^2 \bar{q}_3\right)+4 x \bar{c}^2
   \bar{q}_0 \bar{q}_3 \left(S^2-W^2\right)-4 \bar{q}_0^2\right)}{12 x^2 \bar{q}_3-6}
   \Bigg] \nonumber \coma
\end{align}
and $H(a,b)$ is the root sum function, giving the sum of $b(x)$ over all roots $x$ of $a(x)$.

\subsubsection{Supersonic regime}

For the supersonic regime, the same procedure using Eq.~(\ref{eq:supsonic_q}) yields
\begin{align} \label{eq:rSuper}
\ln(r)= \frac{S^2}{2 W^2} \, \frac{\bar{c} \left(S^2-W^2\right) \left(\bar{c} \left( q_2 u_d - \frac{q_1-1}{2}
   \right) \ln \left(1-\frac{u^2}{S^2 \bar{c}^2}\right)+\frac{\coth ^{-1}\left(\frac{S \bar{c}}{u}\right)
   \left(q_1 u_d - q_2 S^2 \bar{c}^2-q_2 u_d^2-q_0\right)}{S}\right)+G(u)-G(0)}{\left(q_2 S^2
   \bar{c}^2+u_d \left(q_2 u_d-q_1\right)+q_0\right){}^2-S^2 \bar{c}^2 \left(2 q_2
   u_d-q_1+1\right){}^2} \coma
\end{align}
where we omitted the $(d)$ superscripts on the coefficients $q$, and defined
\begin{align}
G(u) \equiv & H\Bigg\{
q_2 \left(u_d-x\right){}^2+q_1 \left(x-u_d\right)+q_0-x,  \\
& \frac{\ln (u-x)}{2 q_2 \left(x-u_d\right)+q_1-1} \Big[
q_0 \left(q_2 \bar{c}^2 \left(S^2+W^2\right)+2 u_d \left(q_2 u_d-q_1\right)\right)+q_2^2 S^2
   W^2 \bar{c}^4+u_d^2 \left(q_1-q_2 u_d\right){}^2+q_0^2 \nonumber \\
   & + \bar{c}^2 \left(q_2 x \left(S^2-W^2\right) \left(-2 q_2 u_d+q_1-1\right)+q_2 S^2 u_d \left(q_2
   u_d-q_1\right)+W^2 \left(-3 q_2^2 u_d^2+\left(3 q_1-4\right) q_2
   u_d-\left(q_1-1\right){}^2\right)\right)
\Big]
   \Bigg\} \nonumber \fin
\end{align}

\fixapj{

\subsubsection{Shock standoff distance}
\label{sec:subsecSO}
The standoff distance $\Delta$ may be found by computing $r_s=(1+\Delta)$ from Eq.~(\ref{eq:rSuper}) with $u=u_d$.
In the strong shock limit, the downstream Mach number is given by $M_0=\myS/g=\myW^2/\myS$, and one obtains (after considerable algebra) \begin{equation}
r_s = \gamma ^{G_+} \left[4 \left(2+\gamma +\gamma^{-1}\right)\right]^{G_-} A_-^{B_-} A_+^{B_+}  \coma
\end{equation}
where
\begin{equation}
A_\pm \equiv {\textstyle\frac{\gamma   (\gamma   (5  \gamma -6)-3\pm C)}{4+\gamma
     (\gamma   (-37+\gamma   (19+5  \gamma ))-15\pm
   4  C)}} \coma
\end{equation}
\begin{equation}
B_\pm\equiv {\textstyle \frac{(-1+\gamma )  (\pm 40-2  C  (-1+\gamma )  (4+\gamma
     (-15+\gamma   (-37+\gamma   (19+5  \gamma ))))\pm
     \gamma   (-266+\gamma   (529+\gamma   (27+\gamma
     (-174+\gamma   (-292+7  \gamma   (19+5
   \gamma )))))))}{C  (32+\gamma   (5+\gamma )  (-11+5
   \gamma ))  (4+\gamma   (-19+\gamma   (10+\gamma )))}}
\coma
\end{equation}
\begin{equation}
C \equiv {\textstyle\sqrt{\gamma  (\gamma  (\gamma  (5 \gamma -56)+58)+48)-7}} \coma
\end{equation}
and
\begin{equation}
G_\pm \equiv {\textstyle -2 \pm \frac{2-2\gamma}{32+\gamma(5+\gamma)
   (-11+5 \gamma)}+\frac{8-38\gamma +22\gamma^2}{4+\gamma  (-19+\gamma  (10+\gamma ))}} \fin
\end{equation}
This roughly gives $\Delta\sim 2/(3g)$. Other fits in the range $1<\gamma<2$ include $\Delta\simeq 0.61 g^{-0.94}$ and $\Delta\simeq 0.37(g-1)^{-0.75}$.

}


\end{document}